\documentclass[english,english,english]{article}
\usepackage[T1]{fontenc}
\usepackage[latin1]{inputenc}
\usepackage{setspace}
\usepackage{a4wide}
\usepackage{babel}
\makeatletter

\providecommand{\LyX}{L\kern-.1667em\lower.25em\hbox{Y}\kern-.125emX\@}

\usepackage[T1]{fontenc}
\usepackage[latin1]{inputenc}
\usepackage{a4wide}
\usepackage{babel}
\makeatletter

\usepackage[T1]{fontenc}
\usepackage[latin1]{inputenc}
\usepackage{a4wide}
\usepackage{babel}


\makeatother
\begin{document}

%\textbf{Submission for \underline{Proceedings of the International Conference on Quantum
%                        Information}, Oviedo, Spain July 13-18, 2002}

{\large \hfill{}}\textbf{\large Can there be quantum correlations 
in a mixture of \hfill{} }{\large \par} 
{\large \hfill{}}\textbf{\large two separable states?
\hfill{} }{\large \par}

\hfill{}Aditi Sen(De)\footnote{%
aditi@univ.gda.pl 
} and Ujjwal Sen\footnote{%
ujjwal@univ.gda.pl 
}\hfill{} 

{\footnotesize \hfill{}}{\footnotesize Institute of
Theoretical Physics and Astrophysics, University of Gda\'{n}sk, 80-952
Gda\'{n}sk, Poland}{\footnotesize \hfill{}} {\footnotesize \par{}}{\footnotesize \par}

\doublespacing{

\begin{abstract}

We use a 
recently proposed measure of quantum correlations (work deficit) to measure the strength 
of the nonlocality  
of an  equal mixture of two bipartite orthogonal but locally 
indistinguishable separable states. This gives supporting evidence
of nonzero value for a separable state for this measure of nonlocality.
We also show that this measure of quantum correlations places a different order on the set of states, than the good
asymptotic measures of entanglement. 
And that such a different order imposed on two states by the work deficit
and any entanglement measure cannot be explained by mixedness alone.

\end{abstract}

%----------------------------------------------------------------

Quantum correlations between separated parties can exhibit quite non-intuitive 
properties. And the usual belief was 
that these non-intuitive properties are due to the entanglement between the systems 
that the parties share. 

It would not have been surprising therefore if even \underline{orthogonal} multipartite states
turned out to be indistinguishable if the sharing parties were allowed to operate only
locally. However it was demonstrated that there exist sets of orthogonal \underline{product}
multipartite states which are indistinguishable if the parties apply only local operations 
and communicate classically (LOCC) \cite{nlwe, UPB1, UPB2}. 
This phenomenon of indistinguishability in the case of a \underline{complete} 
orthogonal product basis has been called `nonlocality without entanglement' \cite{nlwe}.
Further 
it was shown that \underline{any} two orthogonal multipartite states can always be distinguished 
locally irrespective of the entanglement in the states \cite{Walgate1}. Later on it was also 
shown that for two nonorthogonal states, the optimal discrimination protocols 
in the inconclusive as well as in the conclusive cases (in certain ranges) can be
implemented  locally \cite{VSPM, CY}. 

There is another twist to these results. Namely, the three maximally entangled states 
\[
\psi_{1}=\frac{1}{\sqrt{3}}\left(\left|00\right\rangle + \omega \left|11\right\rangle +  \omega^{2} \left|22\right\rangle\right),   
\psi_{2}=\frac{1}{\sqrt{3}}\left(\left|00\right\rangle + \omega^{2} \left|11\right\rangle + \omega \left|22\right\rangle\right),  
\psi_{3}=\frac{1}{\sqrt{3}}\left(\left|01\right\rangle + \left|12\right\rangle + \left|20\right\rangle\right)
\]
(where \(\omega\) is a nonreal cube root of unity)
in \(3 \otimes 3\) are distinguishable locally. But if the third maximally entangled state 
is swapped by the \underline{product} state \(\left|01\right\rangle\), the states
are indistinguishable locally \cite{KPAU}. Therefore, not only is there `nonlocality without
entanglement', there appears to exist `more nonlocality with 
less entanglement' \cite{LH}. All these results seem to imply that
the concept of nonlocality (in the sense of local indistinguishability of orthogonal states) is  
independent of entanglement.

A set of multipartite orthogonal product states which are not distinguishable
locally is clearly nonlocal in some sense. It would 
be interesting to quantify the amount of 
nonlocality of the set \cite{DP}. But since the states in the set are product, 
a measure of entanglement cannot be used. The average entanglement in any 
such set is zero. 

Meanwhile it was demonstrated 
\cite{datahid1, datahid2} that there exists \underline{two} orthogonal separable \underline{mixed} 
states in \(2 \otimes 2\) (\(\rho^{0}\) and \(\rho^{1}\)) which are
indistinguishable locally, where  \(\rho^{0}\) and \(\rho^{1}\) are given by 
\[
{\rho}^{0} = {\frac{1}{2}}P\left[\left|0\right\rangle\frac{1}{\sqrt{2}}\left(\left|0\right\rangle + \left|1\right\rangle\right)\right]
+ {\frac{1}{2}}P\left[\frac{1}{\sqrt{2}}\left(\left|0\right\rangle + \left|1\right\rangle\right)\left|0\right\rangle\right]
\]
and
\[{\rho}^{1} = {\frac{1}{2}}P\left[\left|1\right\rangle\left|1\right\rangle\right]
+ {\frac{1}{2}}P\left[\frac{1}{\sqrt{2}}\left(\left|0\right\rangle - \left|1\right\rangle\right)\frac{1}{\sqrt{2}}\left(\left|0\right\rangle - \left|1\right\rangle\right)\right],
\]
with \(P[\left|\psi \right\rangle] = \left|\psi\right\rangle\left\langle \psi \right|\).
Therefore, intuitively speaking, a mixture of \({\rho}^{0}\) and \({\rho}^{1}\), although 
separable, contains traces of nonlocality. And this leads us to consider, in a
 quantitative way, 
the strength of the nonlocality that Alice and Bob possess when they 
share a mixture of \({\rho}^{0}\) and \({\rho}^{1}\). 
There has been recent works \cite{delta1, delta2} proposing a 
measure of quantum correlations (or nonlocality),
 which has been indicated to be a broader notion than 
just entanglement. In fact, it has been argued there that an unequal mixture of the pure 
product states exhibiting 
nonlocality without entanglement \cite{nlwe} would probably have a nonzero amount of this 
measure of nonlocality. It would therefore be interesting to find whether a mixture 
of \({\rho}^{0}\) and \({\rho}^{1}\) has a nonzero amount of that measure.

In this paper, we  show that an equal mixture of the separable states \(\rho^0\) and \(\rho^1\), has a nonzero 
value for the measure of quantum correlations proposed in Refs. \cite{delta1, delta2}, when 
we restrict to one-way classical communication. This measure of quantum correlations places a different order
on the set of states than the ``good'' asymptotic measures of entanglement. And such different 
order cannot be explained by different amounts of mixedness of the states.

The proposed measure of quantum correlations (called `work deficit')
for any bipartite state \(\rho\) is \cite{delta1}
\[\triangle (\rho) \equiv W_{t} - W_{l}.
\]
\[W_{t}(\rho) = n - S(\rho)
\]
is defined as the amount of `work' that can be obtained by operations on the whole system. Here
 \(n = \log_{2} \left(\dim H\right)\), where H is the Hilbert space on which \(\rho\) is defined and \(S(\rho)\)
is the von Neumann entropy of \(\rho\). On the other hand, \( W_{l}(\rho)\) is defined as the 
amount of `work' that can be obtained if \(\rho\) is acted upon by LOCC. But since one is dealing here
 with entropies, care must be taken so that all entropies transferred via ancillas are accounted
for. To maximize the local work \(W_{l}\), one can for example consider the following strategy. 
Suppose the bipartite state \(\rho\) is shared between Alice and Bob. Alice makes the projection 
measurement on her part of the state \(\rho\), in some orthogonal basis \(\left\{\left|i\right\rangle\right\}\).
And let the state produced at Bob, when Alice's outcome is \(\left |i\right\rangle\), be \( \xi ^{i}\).
That is \(\xi^{i} = P\left[{\left|i\right\rangle}_{A}\right]\otimes I_{B} {\rho}_{AB} P\left[{\left|i\right\rangle}_{A}\right]\otimes I_{B}\),
where \(I_{B}\) is the identity operator of Bob's part of the Hilbert space on which 
\(\rho_{AB}\) is defined. For the outcome \(\left|i\right\rangle\) in Alice's measurement, the 
total state is transformed into \(P\left[{\left|i\right\rangle}_{A}\right]\otimes{\xi}^{i}_{B}\).
We can think of this whole state to be at Bob's side, once Alice has communicated her measurement result to Bob.
We can therefore be sure of extracting an amount of work equal to 
\[n - S\left({\frac{1}{n_{A}}}P\left[\left|i\right\rangle\right]\otimes{\xi}^{i}\right)\]
by using the above local protocol, where \(n_{A}\) is the dimension of Alice's Hilbert space. 
One can think of other strategies and we refer the reader to Refs. \cite{delta1, delta2} 
for a more detailed description.

As we have already noted, this measure has been proposed to be a broader notion of nonlocality
than just entanglement and it seems that there could exist separable states which produce a 
nonzero value of this measure of nonlocality. A potential candidate for such an effect could be
a mixture of \({\rho}^{0}\) and \({\rho}^{1}\). For definiteness, we consider the equal mixture 
of \({\rho}^{0}\) and \({\rho}^{1}\):
\begin{equation}
\label{rho}
\rho = {\frac{1}{2}}{\rho}^{0} + {\frac{1}{2}}{\rho}^{1}.
\end{equation}
We see that \(W_{t}\left(\rho\right) = 2 - 1.81128 = 0.18872\) (upto
5 decimal places). To find \(W_{l}\left(\rho\right)\), one has 
to optimize over all LOCC protocols. If we restrict ourselves to projection measurements on say,
Alice's side (without adding any ancilla) and consider only one-way classical communication (from
Alice to Bob), then the optimization over all such protocols yields 
\[W_{l} = 2 - 1.87852 = 0.12148\] The corresponding optimal \(\triangle\) is \(0.06724\), which is positive.
However one can consider positive operator valued measurements (POVM) (or what is the same,
consider projection measurements after adding an ancilla (cf. \cite{peres})) and there seems to be
no indication as to how many outcomes should be considered. One may also consider protocols with two-way 
classical communication. We just remark here that the structure of the state \(\rho\) may lead 
one to believe that a POVM on the states \(\left|0\right\rangle\), \(\left|1\right\rangle\),
\((1/\sqrt{2})\left(\left|0\right\rangle + \left|1\right\rangle\right)\) and 
\((1/\sqrt{2})\left(\left|0\right\rangle - \left|1\right\rangle\right)\) would be the 
best POVM for extracting the highest local work \(W_{l}\) (and hence optimal \(\triangle\)). 
However this measurement (supplemented by classical communication) surprisingly yields a lower value of 
\(W_{l} = 0.09215\) than the best projection measurement (with one-way classical communication). 
This seems to indicate that projection measurements produce the best value for \(\triangle\) 
when we restrict ourselves to one-way communication. This therefore supports the conjecture made in 
Refs. \cite{delta1, delta2} that there exist separable states which exhibit some form of nonlocality, by 
producing a nonzero value of \(\triangle\) \cite{amader}.

If the value of work deficit \(\triangle\) is indeed nonzero for the state \(\rho\) given
in eq. (\ref{rho}),
one can make interesting comparisons with this measure of nonlocality with 
other entanglement measures. 

There is an interesting work by Munro et al. \cite{Munro}
trying to find a reason for the different ordering being imposed on states
by the entanglement of formation \cite{huge} and the (maximal) amount of 
violation of Bell inequality \cite{Bell, CHSH, HHH} (see also \cite{VP} in this regard). 
The demonstration of Werner \cite{Werner} that among mixed states
there are ones which are entangled and yet do not violate any Bell inequality,
along with the nonexistence of such a phenomenon for bipartite pure states \cite{Gisin}
(see \cite{Zukowski} however) seems to indicate that  mixedness could 
explain this anomaly. This intuition has however the following problem \cite{MvsBV}:
there are states \(\rho_1\) and \(\rho_2\)
such that keeping their entanglement of formation (\(E_F\)) equal, 
\[
E_F(\rho_1) = E_F(\rho_2),
\]
but with  
\[B(\rho_1) > B(\rho_2)\]
(\(B\) being the amount of violation of Bell inequality), one can have both
\[S(\rho_1) > S(\rho_2)\]
as well as 
\[S(\rho_1) < S(\rho_2),\]
where \(S\) is either the von Neumann or the linearised entropy \cite{BV}. 
Further results were obtained in Ref. \cite{VW}.

If the value of the work deficit \(\triangle\) is nonzero for the state \(\rho\) 
(of eq. (\ref{rho})),
it is possible to make such an exercise to see the role played by mixedness in 
an ordering of states by \(\triangle\) and the measures of entanglement. 

Let us take
\[\rho_1 = P[a\left|00\right\rangle + \left|11\right\rangle];
 \quad \rho_2 = \rho
\]
where \(ab \ne 0
\).

As \(\rho_1\) is always entangled while \(\rho_2\) is a separable state, we have 
\begin{equation}
\label{chukki}
E(\rho_1) > E(\rho_2),
\end{equation}
with respect to any measure of entanglement \(E\).
Suppose now that the work deficit \(\triangle\) for the state \(\rho_2 = \rho\)
is nonzero, as we had tried to argue in this paper. 

Now for pure states the work deficit is exactly equal to the 
unique asymptotic measure of entanglement for pure states \cite{delta1}. And for 
the class \(a\left|00\right\rangle +  b \left|11\right\rangle\), this entanglement
(and therefore the work deficit) ranges continuously from \(0\) to \(1\). Therefore
there are different examples of the pair \(\left\{\rho_1, \rho_2\right\}\)
(for different values of \(a\) and \(b\)), for which
\[\triangle(\rho_1) > \triangle(\rho_2)\]
as well as 
\[\triangle(\rho_1) < \triangle(\rho_2)\]
holds.

But \(\rho_1\) has zero mixedness and so 
\[S(\rho_1) < S(\rho_2)\]
with respect to any measure of mixedness.

Hence a different order between two states as given by their work deficits and the value of any
entanglement measure cannot be explained by their different amounts of mixedness. 

Note that the preceeding discussion cannot hold for separable states which does not 
have a nonzero value of the work deficit.

\textbf{Note 1:} Note here that for eq. (\ref{chukki}) to be true, the entanglement \(E\) 
%in eq. (\ref{chukki}) 
can 
be any measure of entanglement, asymptotic or non-asymptotic.

\textbf{Note 2:} The above discussion shows, somewhat surprisingly, that 
the work deficit of an entangled state can be sometimes \underline{smaller}
than the work deficit for a \underline{separable} state.

The previous discussion shows that in a  \underline{specific} case, the order
of entanglement between two states, does not imply any definite order among their work deficit.
And the considerations were essentially of a restricted nature due to the fact that we were able to 
consider the work deficit only in the situation where one-way classical communication is allowed (see the note added
at the end). However we will now show that such a consideration can be made generic by 
extending the arguments in Ref. \cite{VP}, even in the case of asymptotic work deficit under two-way classical 
communication (two-way work deficit).

Suppose that 
\begin{equation}
\label{Shashank}
{\cal E}(\varrho_1) \leq {\cal E}(\varrho_2) \Leftrightarrow \triangle (\varrho_1) \leq  \triangle (\varrho_2)
\end{equation}
is true for arbitrary states \(\varrho_1\) and \(\varrho_2\). Here \({\cal E}\) 
denotes any measure of entanglement which is defined for all states and 
reduces to von Neumann entropy 
of the single-party reduced density matrix for pure states.
In particular, \({\cal E}\) can be any ``good'' asymptotic measure of entanglement 
(see for example \cite{DHR}). 
%In particular, it  
And \(\triangle\) is now the asymptotic
two-way work deficit (where even POVMs are considered in the local measurements). 
Then following the argument in Ref. \cite{VP} (and remembering the fact that 
asymptotic two-way work deficit is equal to von Neumann entropy of local density martices, in the case of pure 
states \cite{delta2}), one obtains that the condition in eq. (\ref{Shashank}) for arbitrary 
states \(\varrho_1\) and \(\varrho_2\), implies 
that (and also is implied by)
\begin{equation}
\label{S1}
{\cal E}(\varrho) = \triangle(\varrho),
\end{equation}
for all states \(\varrho\).
For mixed states, work deficit is  potentially a different measure of quantum correlations than the 
measures of ``entanglement''. In fact, in this paper, we have tried to argue for this conjecture. 
In any case, there are good asymptotic measures of entanglement, which differ for mixed states, although they 
coincide on pure states. For example, distillable entanglement and entanglement cost are provably different for 
certain states \cite{ECED}. Two-way work deficit cannot of course be equal to both of them. So there are examples 
of \(\varrho\) and \({\cal E}\) for which the relation in eq. (\ref{S1}) cannot hold. Correspondingly, there will exist
examples of pairs, \(\{\varrho_1, \varrho_2\}\), for which the relation in eq. (\ref{Shashank}).

%What we obtain here is that such difference will lead to a different ordering on the 
%set of states, as placed by  any asymptotically good measure 
%of entanglement and work deficit. 

To conclude, we have discussed on the possible nonzero value
of a recently proposed measure of quantum correlations (work deficit)
for an equal mixture of two separable states. These separable 
states are orthogonal (mixed) states (in \(2 \otimes 2\))
and yet they are  locally indistinguishable. The discussion gives supportive 
evidence to 
the conjecture that there exist separable states which possess a nonzero amount
of nonlocality \cite{amader}. We also show that a different order is imposed on the set of states
by ``good'' asymptotic measures of entanglement and work deficit. And such 
different order
% on the states given by the work deficit
%and any entanglement measure 
cannot be explained by the different amounts of mixedness in the states.

\textbf{Note added:}

 After completing this work, we have shown \cite{hugepreparation} that the amount of 
work deficit with one-way classical communication (one-way work deficit)
 for mixtures of Bell states (in \(2 \otimes 2\)) is 
additive. Therefore the asymptotic one-way work deficit is equal to the single-copy one-way work deficit for such states. 
Furthermore, the optimal value of one-way work deficit for mixtures of Bell states, is 
attained for projection-valued measurements (applied only on the system, i.e., ancillas are not required).
The equal mixture of two separable states  \(\rho\) (of eq. (\ref{rho})) considered in this paper,
is a mixture of Bell states upto local unitary transformations. Work deficit (one-way or two-way)
is invariant under local unitary transformations. Therefore the value of one-way work deficit 
(\(\approx 0.06724\)) for the 
state \(\rho\), obtained in this paper 
by considering only a single copy of the state and optimizing over 
projection measurements only, 
is  actually the  asymptotic work deficit   
by one-way classical communication (where POVMs are also included in the measurement before 
the classical communication).

\textbf{Acknowledgements}

We acknowledge useful discussions with David. P. DiVincenzo and Barbara M. Terhal 
 (during the European Research Conference on Quantum Information at 
San Feliu de Guixols, Spain, March 2002),
Sibasish Ghosh, Karol Horodecki, Micha{\l} Horodecki,
Pawe{\l}  Horodecki, Ryszard 
Horodecki and Jonathan Oppenheim. 
This work is supported by the EU Project
EQUIP Contract No. IST-1999-11053 and by the University of Gda\'{n}sk, 
Grant No. BW/5400-5-0236-2. 

}

\begin{thebibliography}{10}
\bibitem{nlwe}Bennett, C.H., DiVincenzo, D.P.,  Fuchs, C.A.,  Mor, T., Rains, E., 
Shor, P.W.,  Smolin, J.A., and  Wootters, W.K., 1999, \underline{Physical Review A},
\textbf{59}, 1070-1091.
 \bibitem{UPB1}Bennett, C.H. ,  DiVincenzo, D.P., 
 Mor, T.,  Shor, P.W.,  Smolin,  J.A., and  Terhal, B.M.,  1999, 
\underline{Physical Review Letters}, 
 \textbf{82}, 5385-5388.
\bibitem{UPB2}DiVincenzo, D.P., Mor, T., Shor, 
P.W.,
 Smolin, J.A., and  Terhal, B.M., http://xxx.lanl.gov/abs/quant-ph/9908070.
\bibitem{Walgate1} Walgate, J.,  Short, A.J., Hardy., L., and  Vedral, V., 2000,
 \underline{Physical Review Letters}, \textbf{85}, 4972-4975.
\bibitem{VSPM} Virmani, S.,  Sacchi, M.F.,  Plenio, M.B., and Markham, D.,
 2001, \underline{Physics Letters A}, \textbf{288}, 62 
\bibitem{CY} Chen, Y.-X., and  Yang, D., 2001, \underline{Physical Review A}, \textbf{64}, 
064303:1-3; 
 Chen,Y.-X., and  Yang, D., 2002,
\underline{Physical Review A }, \textbf{65}, 022320:1-4.
\bibitem{KPAU} Horodecki, M., Sen(De), A.,  Sen, U., and Horodecki, K., 2003,
 \underline{Physical Review Letters},  \textbf{90}, 047902:1-4.
%now at http://xxx.lanl.gov/abs/quant-ph/0204116.
\bibitem{LH}This was coined by L. Hardy during a discussion
at the 
International Conference on Quantum
                        Information, Oviedo, Spain, July 2002.
\bibitem{DP} DiVincenzo, D.P., private communication.
\bibitem{datahid1}DiVincenzo, D.P., Leung,  D.W., and Terhal, B.M., 
http://xxx.lanl.gov/abs/quant-ph/0103098.
\bibitem{datahid2}Terhal, B.M.,  DiVincenzo,  D.P., and   Leung, D.W., 2001,\underline{ 
Physical Review Letters},
\textbf{86}, 5807-5810. 
\bibitem{delta1}Oppenheim, J.,  Horodecki, M., Horodecki, P., and  Horodecki, R., 
2002, \underline{Physical Review Letters}, \textbf{89}, 180402:1-4.
%, now at 
%http://xxx.lanl.gov/abs/quant-ph/0112074.
\bibitem{delta2}Horodecki, M., Horodecki, K., Horodecki, P.,
Horodecki, R., Oppenheim, J., Sen(De), A., and  Sen, U., 2003, 
\underline{Physical Review Letters},  \textbf{90}, 100402:1-4.
% now at 
%http://xxx.lanl.gov/abs/quant-ph/0207168.
\bibitem{peres}Peres, A., 1993, \underline{Quantum Theory: Concepts and Methods} (Kluwer 
Academic Publishers).

\bibitem{amader} See note added at the end of the paper.
\bibitem{Munro}Munro, W.J. ,  Nemoto, K., and  White, A.G., 2001, \underline{Journal of Modern Optics},
 \textbf{48}, 1239-1246. 
\bibitem{huge}Bennett,  C.H. , DiVincenzo, D.P., Smolin, J.A., and 
 Wootters, W.K., 1996, \underline{Physical Review A}, \textbf {54}, 3824-3851. 
\bibitem{Bell} Bell, J.S., 1964, \underline{Physics}, \textbf{1}, 195. 
\bibitem{CHSH} Clauser, J.,  Horne, M., Shimony, A., and  Holt, R., 1969, 
\underline{Physical Review Letters},
 \textbf{23}, 880-883.
\bibitem{HHH} Horodecki, R., Horodecki, P., and Horodecki, M.,  
1995, \underline{Physics Letters A }, \textbf{200}, 340-344. 
\bibitem{VP} Virmani, S., and Plenio, M.B., 2000, \underline{Physics Letters A}, \textbf{268}, 
31-34. 
\bibitem{Werner} Werner, R.F., 1989, \underline{Physical Review A}, \textbf{40}, 4277-4281.
\bibitem{Gisin} Gisin, N., 1991, \underline{Physics Letters A}, \textbf{154}, 201; 
 Gisin, N., and  Peres, A., 1992, \underline{Physics Letters A}, \textbf{162}, 15. 
\bibitem{Zukowski} Scarani, V., and Gisin, N., 2001, \underline{Journal of Physics A},
 \textbf{34}, 6043;  \.Zukowski, M., Brukner, {\v C.}, Laskowski, W., and  Wie{\' s}niak, M.,
2002, \underline{Physical Review Letters}, \textbf{88}, 210402:1-4 ;  Sen(De), A., Sen, U.,
 and  \.Zukowski, M., 2002, \underline{Physical Review A}, \textbf{66}, 062318:1-4.
\bibitem{MvsBV} Ghosh, S., Kar, G., Sen(De), A., and  Sen, U., 2001,
 \underline{Physical Review A}, \textbf{64}, 044301:1-3.
\bibitem{BV} Bose, S., and  Vedral, V., 2000, \underline{Physical Review A},  
\textbf{61}, 040101:1-2.
\bibitem{VW}Verstraete, F., and  Wolf, M. M., http://xxx.lanl.gov/abs/quant-ph/0112012.
\bibitem{DHR}Horodecki, M., \underline{Quantum Information and Computation}, 2001, \textbf{1}, 3;
Donald, M.J., Horodecki, M., and Rudolph, O., 2002, \underline{Journal of Mathematical 
Physics}, \textbf{43},     4252-4272.
\bibitem{ECED}Vidal, G., and  Cirac, J.I., 2001, \underline{Physical  Review Letters},
\textbf{86}, 5803-5806; Vidal, G.,
and Cirac, J.I., 2002, \underline{Physical Review A} \textbf{65}, 012323; Vidal, G.,  Dur, W., and Cirac,
J.I., 2002, \underline{Physical Review Letters} \textbf{89}, 027901; 
Horodecki, M.,  Sen(De), A., and  Sen, U., 
http://xxx.lanl.gov/abs/quant-ph/0207031;  Vollbrecht, K.G.H., Werner, R.F., and Wolf, M.M., 
http://xxx.lanl.gov/abs/quant-ph/0301072.
\bibitem{hugepreparation} Horodecki, M., Horodecki, P.,  Oppenheim, J.,  Sen(De), A.,  Sen, U.,
 and Synak, B., in preparation.

\end{thebibliography}
\end{document}